%% file: Rxrx3-core.tex
\documentclass{article} 
\usepackage{iclr2025_conference,times}
\lmrlfinalcopy
\input{math_commands.tex}

\usepackage{hyperref}
\usepackage{url}

\usepackage{array}
\usepackage{booktabs}
\usepackage{float}
\usepackage{graphicx}
\usepackage{adjustbox}
\usepackage{authblk}

\title{RxRx3-core: Benchmarking drug-target \\ interactions in High-Content Microscopy}

\author[1]{Oren Kraus}
\author[1]{Federico Comitani}
\author[1]{John Urbanik}
\author[1]{Kian Kenyon-Dean}
\author[1]{Lakshmanan Arumugam}
\author[1]{Saber Saberian}
\author[2]{Cas Wognum}
\author[1]{Safiye Celik}
\author[1]{Imran S. Haque}

\affil[1]{Recursion\\\texttt{\{first.last\}@recursion.com}}
\affil[2]{Valence Labs}

\begin{document}

\maketitle

\begin{abstract}
High Content Screening (HCS) microscopy datasets have transformed the ability to profile cellular responses to genetic and chemical perturbations, enabling cell-based inference of drug-target and gene-gene interactions. However, the adoption of representation learning methods for HCS data has been hindered by the lack of accessible datasets and robust benchmarks. To address this gap, we present \textbf{RxRx3-core}, a curated and compressed subset of the RxRx3 dataset, and associated drug-target and gene-gene interaction benchmarks. At just 18GB, RxRx3-core significantly reduces the size barrier associated with large-scale HCS datasets while preserving critical data necessary for benchmarking representation learning models against zero-shot perturbation interaction prediction tasks. RxRx3-core includes 222,601 microscopy images spanning 736 CRISPR knockouts and 1,674 compounds at 8 concentrations. RxRx3-core is available on HuggingFace and Polaris, along with pre-trained embeddings and benchmarking code, ensuring accessibility for the research community. By providing a compact dataset and robust benchmarks, we aim to accelerate innovation in representation learning methods for HCS data and support the discovery of novel biological insights.
\\

Dataset: \href{https://huggingface.co/datasets/recursionpharma/rxrx3-core}
{https://huggingface.co/datasets/recursionpharma/rxrx3-core} \\
Polaris Platform: \href{https://polarishub.io/datasets/recursion/rxrx3-core}
{https://polarishub.io/datasets/recursion/rxrx3-core} \\
Benchmarks: \href{https://github.com/recursionpharma/EFAAR_benchmarking} {https://github.com/recursionpharma/EFAAR\_benchmarking} \\
For more information about \textbf{RxRx3-core} please visit \href{https://www.rxrx.ai/rxrx3-core}{rxrx.ai/rxrx3-core}
\end{abstract}

\section{Introduction}

Understanding and quantifying how cells respond to genetic and chemical perturbations — and relating these responses — is a central challenge in biological research ~\citep{PrzybylaNatGeneticsReviews2022,VincentPhenotypic2022}. Advances in automated imaging platforms have transformed our ability to investigate cellular phenotypes triggered by a wide range of perturbations ~\citep{BoutrosHCS2015}. High-Content Screening (HCS) systems, which integrate automated microscopy with robotic liquid handling, enable systematic profiling of cellular responses on an unprecedented scale. Publicly available HCS datasets, such as RxRx3 ~\citep{fay2023rxrx3} and JUMP ~\citep{chandrasekaran2023jump}, exemplify this scalability, including millions of cellular images spanning hundreds of thousands of unique perturbations.

\begin{figure}[t]
    \centering
    \includegraphics[width=\textwidth]{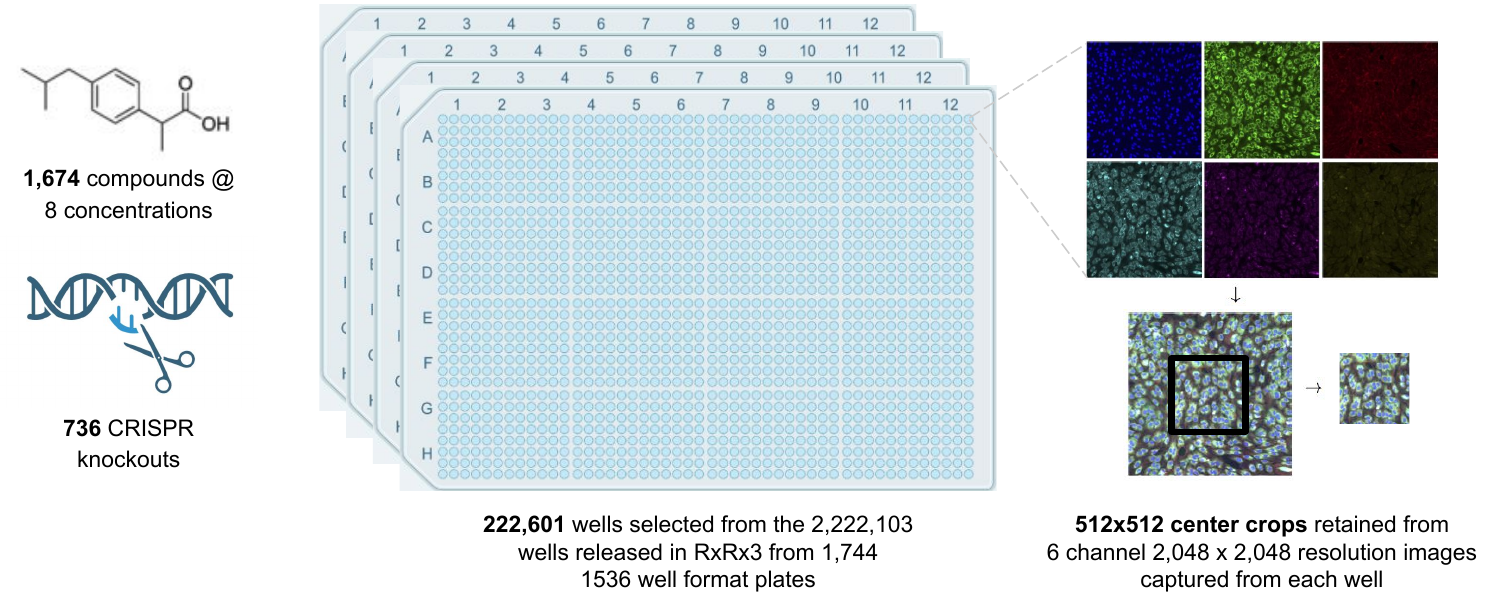}
    \caption{Overview of RxRx3-core, a subset of RxRx3 \citep{fay2023rxrx3}, designed for benchmarking representation learning models.}
    \label{fig:rxrx3-core_overview}
\end{figure}

While the vast scope of HCS datasets creates opportunities to uncover novel biological insights, it also introduces technical challenges in deriving meaningful and interpretable features from these images. Traditionally, HCS images are processed using pipelines that rely on cell segmentation, feature extraction, and subsequent analyses tailored to specific datasets ~\citep{10.1038/nmeth.4397}. Although these workflows have enabled significant discoveries ~\citep{BoutrosHCS2015}, they often require complex and resource-intensive optimization, through proprietary software or open-source tools ~\citep{carpenter2006cellprofiler,10.1186/s12859-021-04344-9}. More recently, representation learning methods leveraging deep learning to learn feature representations directly from pixel values in HCS images have emerged as a promising alternative.

Despite their potential, the progress of representation learning in HCS has been limited by the scarcity of accessible and standardized datasets and benchmarks. Existing benchmarks largely focus on straightforward classification tasks, such as identifying cell cycle stages \citep{BBBC048}, protein localization patterns \citep{hpa_localization}, or mechanisms of action~\citep{BBBC021}. However, these tasks fail to represent the complexity of genome-wide relationships that researchers often need to explore. Although recent large-scale datasets like RxRx3 and JUMP became available in 2023, their sheer size — over 100 TB each — along with the lack of standardized benchmarks, has impeded their widespread adoption. 

Previous efforts to release smaller datasets specifically for benchmarking zero-shot prediction of drug-target interactions (DTI) based on compound target annotations, like CPJUMP1 \citep{cpjump1}, have suffered from plate layout confounders, since well positions were not randomized between technical replicates. More recently, supervised learning has been applied to the CPJUMP1 dataset by treating the DTI task as a binary classification of gene-compound pairs \citep{CPJUMP1_supervised}. However this approach was limited by the small number of perturbations in the dataset (302 compounds and 160 genes) and struggled to generalize to unseen genes. The Motive dataset \citep{broad_motive} paired CellProfiler features extracted from the JUMP Cell Painting dataset~\citep{chandrasekaran2023jump} covering 11,000 genes and 3,600 compounds with annotations aggregated from several public databases. This task, however, is framed as a graph learning task leveraging the extracted features rather than a benchmark designed to assess microscopy representation learning models. Benchmarking DTI prediction as a link prediction task introduces several issues including introducing false positives for annotated interactions in which one or both perturbations do not induce a cellular phenotype and the requirement to carefully design train/test splits to ensure model generalization is accurately assessed. Evaluating the proposed benchmarking task with an alternative model to CellProfiler is burdensome as it would require downloading and running inference on $>$15TB of imaging data and subsequently evaluating graph learning models on these features. 

To address these limitations, we introduce \textbf{RxRx3-core}, a dataset specifically designed to facilitate research in representation learning for HCS data. At only 18 GB (17.5GB of images and 0.5Gb of pre-computed embeddings), RxRx3-core is significantly more accessible than its predecessors and is accompanied by benchmarks evaluating perturbation signal magnitude and zero-shot DTI and gene-gene interaction prediction. By providing both a manageable dataset and well-defined benchmarks, we aim to accelerate innovation in representation learning methods for high-content phenotypic screening.

\section{RxRx3-core Dataset}

RxRx3-core is a compressed and accessible subset of RxRx3~\citep{fay2023rxrx3}, a dataset initially released by Recursion in 2023. RxRx3 contains 6-channel fluorescent microscopy images of HUVEC cells stained using a modified Cell Painting protocol ~\citep{10.1038/nprot.2016.105}. The full dataset comprises 2,222,103 experimental wells, encompassing 17,063 CRISPR knockouts and 1,674 commercially available bioactive compounds tested at 8 concentrations each. Most CRISPR knockouts are represented by six unique guide RNAs targeting the gene.

RxRx3-core, outlined in figure \ref{fig:rxrx3-core_overview}, focuses on a smaller, more practical subset, including images from the 736 genes knockouts not blinded in RxRx3, all 1,674 compounds at all 8 concentrations, and relevant control wells. This curated subset retains the critical data necessary for benchmarking representation learning models while substantially reducing dataset size. The key statistics for RxRx3-core are summarized in Table \ref{tab:rxrx3_summary}.

\begin{table}[t]
\caption{RxRx3-core summary statistics}
\centering
\begin{tabular}{@{}l@{\hskip 1cm}r@{}}
\toprule
\textbf{Metric}                          & \textbf{Count}           \\ \midrule
Images (individual fluorescent channel JP2 files) & 1,335,606         \\
Wells                                    & 222,601           \\
Plates sampled                           & 1,744             \\
Unique compounds                         & 1,674             \\
Unique compounds-concentration pairs     & 13,358            \\
Unique CRISPR gene targets               & 736               \\
Unique CRISPR guide RNAs                 & 4,433             \\ 
Unique CRISPR guide RNAs per gene target                 & 6             \\ 
\# of well reps per  CRISPR guide RNA                 & 18            \\ 
\# of well reps per  compounds-concentration pairs      & 4             \\ 
\bottomrule
\end{tabular}
\label{tab:rxrx3_summary}
\end{table}

\subsection{Experimental Details}
\textbf{CRISPR Knockout Experiments:} RxRx3-core includes data from gene001 to gene176 experiments in RxRx3. Each experiment includes 9 repeats of the same subset of perturbations but with randomized well layouts per plate. Each plate contains a combination of control and target guide wells, with only wells that pass quality control filters retained \citep{Celik2022.12.09.519400}. Wells with annotations for 736 genes and a subset of control wells are included.

\textbf{Compound Experiments:} The dataset includes 1,674 compounds tested at 8 concentrations. Each compound experiment is typically repeated four times, with control compounds tested more frequently. As with CRISPR experiments, only wells that pass quality control filters \citep{Celik2022.12.09.519400} have been included in RxRx3-core.

\textbf{Compressing RxRx3-core:} The original RxRx3 dataset exceeded 100 TB and the majority of its well treatment metadata was blinded, making it challenging for widespread use within the research community. To address this limitation, we compressed the dataset to just 17.5 GB while preserving its utility for benchmarking. The data preparation steps applied are outlined in Table \ref{tab:reduction_factors}. These compression steps reduced the image set size for RxRx3-core from over 10 TB for the set of unblinded wells included in RxRx3-core to 17.5 GB without compromising the ability to distinguish representation learning methods effectively.

\begin{table}[t]
\caption{Data processing steps to create RxRx3-core.}
\centering
\begin{tabular}{@{}l@{\hskip 1cm}r@{}}
\toprule
\textbf{Data Preparation Step}                           & \textbf{Compression Factor} \\ \midrule
Subset to unblinded wells in RxRx3 & 1/10 \\
Included 512x512 center crops from original 2048x2048 images & 1/16                      \\
Converted image format from uint16 to uint8              & 1/2                       \\
Applied JPEG 2000 compression using quality layers [80, 40, 20] & 1/16                      \\ \bottomrule
\end{tabular}
\label{tab:reduction_factors}
\end{table}

\textbf{Accessibility of RxRx3-core:} RxRx3-core is designed to be highly accessible to the machine learning community. The dataset is available on HuggingFace (\href{https://huggingface.co/datasets/recursionpharma/rxrx3-core}{huggingface.co/datasets/recursionpharma/rxrx3-core}) and can be loaded directly using the HuggingFace \texttt{datasets.load\_dataset} function. Additionally, embeddings generated by the OpenPhenom\mbox{-}S/16 \citep{kraus2024masked} model are provided on the dataset page. The dataset is also available on the Polaris platform \citep{polaris_NMI,polaris_zenodo} for drug discovery (\href{https://polarishub.io/datasets/recursion/rxrx3-core}{polarishub.io/datasets/recursion/rxrx3-core}).

To facilitate benchmarking, we have released the benchmarking code in the EFAAR repository(\href{https://github.com/recursionpharma/EFAAR_benchmarking} {github.com/recursionpharma/EFAAR\_benchmarking}) and on the Polaris platform  (\href{https://polarishub.io/benchmarks/recursion/rxrx-compound-gene-activity-benchmark}{polarishub.io/benchmarks/recursion/rxrx-compound-gene-activity-benchmark}), enabling a streamlined model evaluation and development.

\section{Phenomic representation learning}

On the RxRx3-core images, we evaluated a manual feature extraction method (CellProfiler) compared to three self-supervised vision transformer-based foundation models trained as masked autoencoders (MAE), as summarized in Table \ref{tab:representation_learning}. Together, the four representation learning methods we evaluate provide a strong comparison between open and proprietary approaches for feature extraction in high-content phenotypic screening.

\textbf{CellProfiler} is a widely used open-source tool for cell segmentation and feature extraction. CellProfiler features for RxRx3-core were computed with version 2.2.0 \citep{kamentsky2011improved}. Single cells were segmented after applying illumination correction and shape, intensity, and texture features were extracted from each cell, resulting in 952 dimensional profiles. 

\textbf{OpenPhenom-S/16} is a fully open-access model trained on publicly available datasets and is available on HuggingFace (\href{https://huggingface.co/recursionpharma/OpenPhenom}{huggingface.co/recursionpharma/OpenPhenom}). It leverages a channel-agnostic MAE architecture \citep{kraus2024masked} with a 25 million parameter ViT-S/16 encoder backbone and 6 decoders (one for each channel). It was trained for 100 epochs on RxRx3 \citep{fay2023rxrx3} and cpg0016 (CRISPR and ORF subsets from sources 4 and 13 of JUMP) \citep{chandrasekaran2023jump}. Being channel-agnostic, the model tokenizes each image channel's 16x16 pixel patches separately, yielding 1536 tokens for a 6-channel image at inference time and 384 unmasked tokens visible during its training with a 75\% mask ratio.

\textbf{Phenom-1} is a standard MAE \citep{he2022masked} with a ViT-L/8 encoder backbone trained with a 75\% mask ratio \citep{kraus2024masked}. It is a proprietary model trained on over 3.5 billion image crops sampled from the RPI-93M dataset, which contains 93 million unique wells across a large variety of perturbations and cell types. Its 8x8x6 pixel patch size means that each image crop is tokenized to 1024 tokens, and during training the encoder processes 256 tokens to predict the rest.

\textbf{Phenom-2} is a standard MAE with a ViT-G/8 encoder backbone trained with a 75\% mask ratio \citep{vitally_consistent}. We take its embeddings from the \textit{trimmed} layer, where linear probes found that the best representations from this model were at the intermediate 38th layer of the encoder, rather than the final 48th layer. This proprietary model was trained on over 8 billion microscopy image crops sampled from Phenoprints-16M, a specially curated dataset of 16 million wells with a diverse set of meaningful perturbations. Both Phenom-1 and Phenom-2 include non-public 6-channel brightfield and CellPainting images in their training datasets.

\begin{table}[t]
\caption{Representation learning methods examined in this work.}
\centering
\begin{tabular}{@{}l@{\hskip 1cm}l@{\hskip 1cm}l@{\hskip 1cm}r@{}}
\toprule
\textbf{Model} & \textbf{Training Data}     & \textbf{Architecture} & \textbf{Parameters} \\ \midrule
CellProfiler                    & -                          & -                     & -                    \\
OpenPhenom-S/16                      & RxRx3 + cpg0016            & CA-MAE ViT-S/16       & 25M                 \\
Phenom-1                             & RPI-93M                   & MAE ViT-L/8           & 307M                \\
Phenom-2                             & Phenoprints-16M           & MAE ViT-G/8           & 1,860M              \\ \bottomrule
\end{tabular}
\label{tab:representation_learning}
\end{table}

\section{Results}

RxRx3-core enables the computation of benchmarks that evaluate the quality of perturbation embeddings in capturing biological signals across key tasks: distinguishing perturbations from control populations and zero-shot retrieval of drug-target and gene-gene interactions. Before calculating these benchmarks, crop aggregation and batch alignment techniques are applied to phenomics embeddings from representation learning models. First, the four tiled embeddings from every 256x256x6 crop of each 512x512x6 microscopy image are mean-aggregated. Next, a crucial batch alignment step is used to center the latent space on control samples and align experimental batches, ensuring comparability across conditions \citep{Celik2022.12.09.519400}.

We use a PCA-based alignment technique, PCA-CS (Principal Component Analysis with Centering and Scaling), to achieve this. First, a PCA transformation is fitted on all control samples (without reducing dimensionality) and applied to all data points. Next, for each experimental batch, embeddings are centered and scaled relative to the corresponding batch controls. These pre-processing steps standardize the embeddings and minimize batch effects. Following alignment, we compute two benchmarks to assess the embeddings' information content.

\subsection{Perturbation signal benchmark}

The energy distance \citep{energy_distance} quantifies the separation between the distribution of replicate embeddings for a perturbation and that of negative controls, effectively serving as a measure of perturbation effect size in a high-dimensional space. To evaluate this, we calculate the distance between the distribution of embeddings for the replicate perturbation embeddings and the embeddings of the control units, employing statistical methods based on energy metrics.

Assuming access to two sets of embeddings $\mathbf{x}_1, \dots, \mathbf{x}_{n_1}$ (representing query perturbation units) and $\mathbf{y}_1, \dots, \mathbf{y}_{n_2}$ (representing negative control units), the energy distance is defined as
\[
\text{energy}_g = \frac{2}{n_1 n_2} \sum_{i=1}^{n_1} \sum_{j=1}^{n_2} \| \boldsymbol{x}_i - \boldsymbol{y}_j \| 
- \frac{1}{n_1^2} \sum_{i=1}^{n_1} \sum_{j=1}^{n_1} \| \boldsymbol{x}_i - \boldsymbol{x}_j \| 
- \frac{1}{n_2^2} \sum_{i=1}^{n_2} \sum_{j=1}^{n_2} \| \boldsymbol{y}_i - \boldsymbol{y}_j \|.
\]

This metric equals zero when the two distributions are identical and increases with greater divergence between them. The benchmark assumes that better representation learning methods will embed more query perturbations farther from control replicates. In this study, we report the mean energy distance (± MAD) across all perturbations, alongside corresponding z-scores, to indicate improvement over a random baseline. In Table \ref{tab:benchmark_performance} we present the results for the perturbation magnitude benchmark using CellProfiler and ViT baselines described above.

\subsection{Drug-target interaction benchmark}

\begin{figure}[t]
    \centering
    \includegraphics[width=\textwidth]{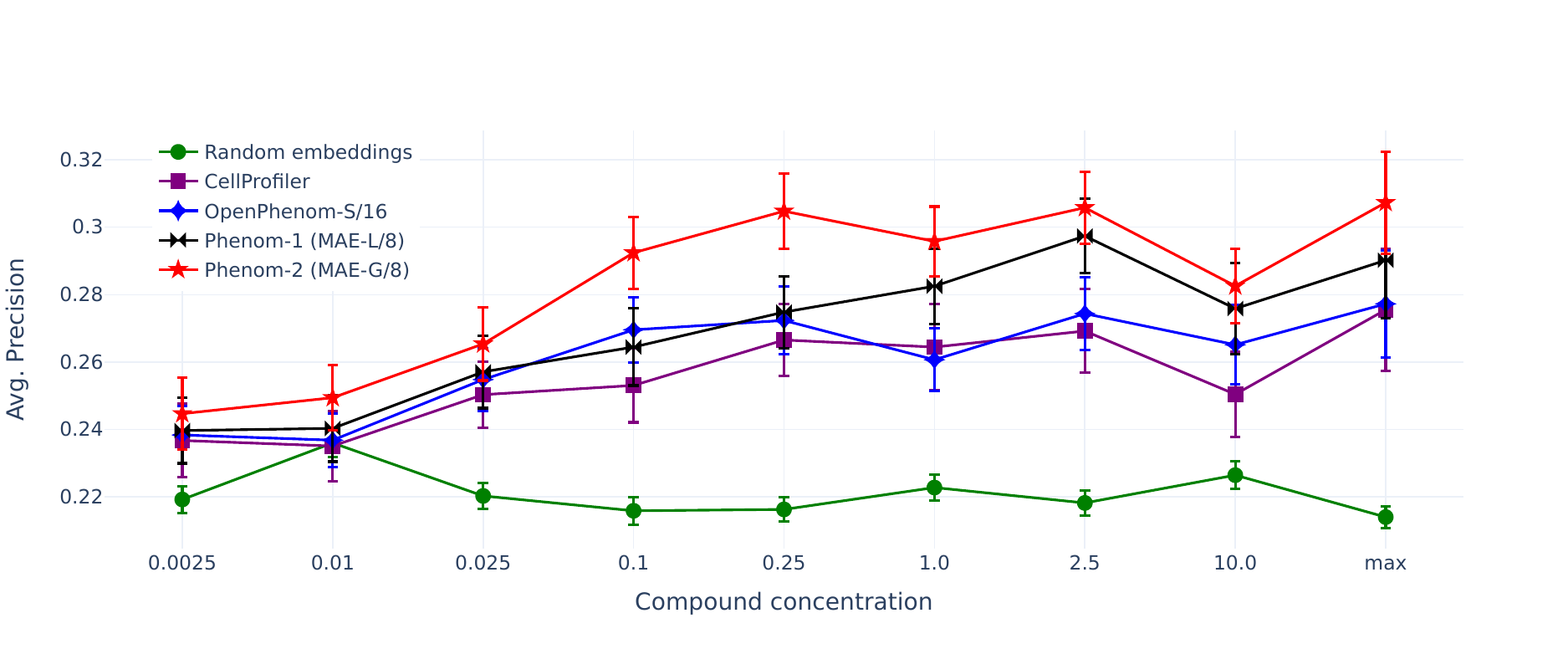}
    \caption{Mean average precision performance on RxRx3-core public benchmark in predicting drug-target interactions, across all compound concentrations with error bars for 100 runs of the benchmark with different random seeds (Table \ref{tab:benchmark_performance}).}
    \label{fig:compound_MAP}
\end{figure}

In addition to reporting perturbation signal magnitude, we also present a benchmark to assess how well representation learning models can relate CRISPR knockouts and compound perturbations in HCS datasets. We evaluate the benchmark for zero-shot prediction of drug-target interactions by analyzing cosine similarities between their embeddings (Figure \ref{fig:compound_MAP}). The compound-gene interaction annotations used in the benchmark were curated from PubChem, Guide to Pharmacology, WIPO, D3R, BindingDB, US Patents, and ChEMBL \citep{bindingdb2007,chembl2024,guide2pharma2024}. To ensure high confidence in our benchmark ground truth, we include only EC50 (IC50) values and avoid mixing measurement types from different assays. These values are aggregated by the median across experiments with the same compound-target pair.  A subset of the resulting drug-target interactions, focusing on key genes, is summarized in Table \ref{tab:compounds_gene}.
This approach assesses whether a model ranks the known target genes of each compound higher than a randomly selected set of genes from the ground truth dataset based on cosine similarity. To quantify confidence, we use the absolute value of the cosine similarity as a proxy, analogous to the probability score in a classifier. For each compound, we calculate the Area Under the ROC Curve (AUC) and average precision as performance metrics.

The final results are summarized by reporting the median AUC and average precision across all compounds, providing a comparison against a random baseline. Given that this precision benchmark requires randomly sampling negative interactions, we report results for 100 random negative sampling seeds of the benchmark with mean and standard deviation. This allows robust comparison against performance of a random baseline via z-score: 

\[ z = \frac{\bar{x}_{\text{model}} - \bar{x}_{\text{baseline}}}{\sqrt{s_{\text{model}}^2 + s_{\text{baseline}}^2}} \]

Table \ref{tab:benchmark_performance} highlights the detailed results along the max axis, which captures the strongest potential interaction for each compound irrespective of its experimental concentration. We note that, while CellProfiler does create salient representations of samples which are more distinguished from controls (Energy), this performance does not necessarily correlate to the primary compound-gene benchmark. These results provide evidence that deep learning features from the Phenom models are better at capturing compound-gene activity in this image data than the CellProfiler features. Overall, Phenom-2's strong performance on each task demonstrates the importance of scaling self-supervised learning to yield biologically meaningful representations, and furthermore motivates continued development of deep learning techniques to continue improving drug discovery.

\begin{table}[t]
\centering
\caption{Performance on the public RxRx3-core drug-target interaction and RxRx3-core perturbation magnitude benchmark, measuring mean average precision ($\pm$ STD over 100 random seeds benchmarking with different negative samples) in predicting compound activity against target genes, and mean energy distance ($\pm$ MAD over all perturbations) separating perturbation embeddings from controls with corresponding z-scores of improvement over a random baseline.}
    \centering
    \begin{adjustbox}{width=\textwidth}
        \begin{tabular}{@{}lccccc@{}}
        \toprule
        \textbf{Method} & \textbf{Average Precision} & \textbf{Comp. z-score ↑}  & \textbf{Energy dist.}    & \textbf{Energy z-score ↑}  \\
            \midrule
        Random embeddings & $0.214 \pm 0.003$ & 0.00 & $0.728 \pm 0.05$ & 0.00 \\
        Cell Profiler & $0.276 \pm 0.018$  & 3.34 & $6.245 \pm 1.32$ & 4.165 \\
        OpenPhenom-S/16 & $0.277 \pm 0.016$ & 3.89 & $3.315 \pm 0.78$  & 3.310 \\
        Phenom-1 & $0.290 \pm 0.017$  & 4.35  & $4.848 \pm 1.07$  & 3.836 \\
        Phenom-2 & $\textbf{0.307} \pm 0.015$  & \textbf{6.04} & $\mathbf{6.636} \pm 1.33$ & \textbf{4.435} \\
            \bottomrule
        \end{tabular}
    \end{adjustbox}
\label{tab:benchmark_performance}
\end{table}

\subsection{Gene-gene interaction benchmark}

In addition to benchmarking drug-target interactions, RxRx3-core can be used to benchmark representation learning models against zero-shot gene-gene interaction prediction. The task, previously described in \cite{Celik2022.12.09.519400} and evaluated in \cite{kraus2024masked} and \cite{vitally_consistent}, leverages embeddings from representation learning models to build biological maps that can be used to uncover novel and meaningful interactions between perturbations. To assess the ability of a map to represent known biological relationships, we employ four annotation sources containing gene-gene interactions as benchmarks (Table \ref{tab:bmdb_datasets}).

\begin{table}[t]
    \caption{Description of datasets used for curating gene-gene interactions.}
    \centering
    \begin{tabular}{ll}  
        \toprule
        \textbf{Dataset} & \textbf{Description} \\
        \midrule
        CORUM \cite{CORUM} & Highly-curated gene clusters of protein complexes \\
        hu.MAP \citep{hu_map} & Comprehensive atlas of interactions and protein complexes \\
        Reactome \citep{REACTOME} & Protein-protein interactions based on curated pathways \\
        StringDB \citep{stringdb} & Proteins associations based on functional interactions \\
        \bottomrule
    \end{tabular}
    \label{tab:bmdb_datasets}
\end{table}

The core hypothesis is that a map capable of accurately capturing known relationships provides a strong representation of biological systems and is likely to reveal novel interactions. To evaluate this, closely following \cite{Celik2022.12.09.519400} Multivariate \textbf{known biological relationship recall} benchmark, we compute recall for the most extreme 10\% of pairwise relationships based on cosine similarity between embeddings. This includes both the top 5\% (high similarity) and bottom 5\% (low similarity) of the pairwise similarity distribution, as the latter may indicate opposing functions between genes; for example, we often find that the gene knockouts for MTOR and TSC2 point in opposite directions with significant negative cosine similarities to each other. A random or uninformative map serves as a baseline, expected to achieve a recall of 10\%.

\begin{table}[t]
    \centering

\caption{Multivariate \textbf{known biological relationship recall} obtained below the 5th and above the 95th percentiles (0.05-0.95 threshold) of all  gene-gene cosine similarities against the null distribution, as per \cite{Celik2022.12.09.519400,kraus2024masked,vitally_consistent}. We report the mean recall over 3 different random seeds of sampling the null distribution in each benchmark run ($\sigma \leq \pm0.001$), computed over four different gene-gene interaction benchmark databases: CORUM, hu.MAP, Reactome-PPI, and StringDB. For \textbf{RxRx3-core} and the \textbf{JUMP-CP (cpg0016)} the embeddings are batch-corrected using a PCA-CS aligner. For \textbf{RxRx3} dataset, the embeddings are transformed with a TVN aligner \citep{TVN2017} followed by a chromosome arm bias correction \cite{lazar2023high}.}
    \begin{tabular}{lcccccc}

            \toprule
            \textbf{Model}           & \textbf{CORUM} & \textbf{hu.MAP} & \textbf{Reactome}  & \textbf{StringDB} \\ 
            \midrule
            \multicolumn{1}{l}{\textbf{RxRx3-core}} \\
            \multicolumn{1}{c}{\textit{736 genes \quad \# relationships}} & \textit{1,209} & \textit{958} & \textit{569} & \textit{1,737} \\
            Random embeddings                     & .107   & .111   & .107   & .115 \\
            CellProfiler                          & .361   & .444   & .160   & .330 \\
            OpenPhenom-S/16                       & .300   & .352   & .158   & .281 \\
            Phenom-1                              & .395   & .482   & .188   & .349 \\
            Phenom-2                              & \textbf{.486}   & \textbf{.553}   & \textbf{.197}   & \textbf{.415} \\
            \midrule
            \multicolumn{1}{l}{\textbf{RxRx3}} \\
            \multicolumn{1}{c}{\textit{17,063 genes \quad \# relationships}} & \textit{29,459} & \textit{47,699} & \textit{14,493} & \textit{43,707} \\
            CellProfiler & .503 & .354 & .190 & .414 \\
            OpenPhenom-S/16 & .549 & .374 & .229 & .429 \\
            Phenom-1 & .609 & .434 & .251 & .472  \\
            Phenom-2 & \textbf{.615} & \textbf{.437} & \textbf{.263} & \textbf{.492}  \\
            \midrule
            \multicolumn{1}{l}{\textbf{JUMP-CP (cpg0016)}} \\ 
            \multicolumn{1}{c}{\textit{7,796 genes \quad \# relationships}} & \textit{8,401} & \textit{9,280} & \textit{8,236} & \textit{17,548} \\       
            CellProfiler &  .219 & .184 & .131 & .191 \\
            OpenPhenom-S/16  &  .233 & .199 & .154 & .214 \\
            Phenom-1 &  .248 & .208 & .160 & .226 \\
            Phenom-2 &  \textbf{.264} & \textbf{.215} & \textbf{.165} & \textbf{.235} \\
    \bottomrule
    \end{tabular}
\label{tab:recall_values}
\end{table}

To calculate recall, we first compute pairwise cosine similarities between the aggregated embeddings of all perturbed genes in RxRx3-core. Self-links, where a gene is compared to itself, are excluded because their cosine similarity is always one, which would bias the results. “Predicted links” are defined as relationships falling within the top or bottom 5\% of the similarity distribution. Recall is then determined for each annotation source as the proportion of true predicted links relative to all possible links. Since datasets differ in the number of genes they include, we adjust recall calculations to consider only gene-gene interactions where both genes are present in the dataset. Therefore, recall results from the subset of 736 gene knockouts (KOs) in RxRx3-core are not directly comparable with other work \citep{kraus2024masked,vitally_consistent} which measures recall across a broader set of 17,000 genes. Despite the subset of genes available in RxRx3-core, Table \ref{tab:recall_values} shows that self-supervised MAEs outperform CellProfiler features as the models are scaled to larger datasets and architectures. 

To demonstrate how models rank on this metric when a larger set of gene-KOs are leveraged, we report the same gene-gene relationship recall benchmark \citep{Celik2022.12.09.519400} on the full RxRx3 dataset, including 17,063 gene-KOs, and on the JUMP dataset (cpg0016) \citep{chandrasekaran2023jump}, including a subset of 7,976 gene-KOs, in Table \ref{tab:recall_values}. As shown in the table, the number of interactions evaluated grows significantly as more CRISPR knockouts are included in the datasets. For post-processing RxRx3-core and JUMP-CP embeddings, we use PCA with center-scaling for standard batch correction alignment. We observe that all MAEs, including OpenPhenom-S/16,  perform better than the CellProfiler feature extraction baseline \citep{carpenter2006cellprofiler,kamentsky2011improved} when a larger set of genes and gene-gene interactions are evaluated. For post-processing the full RxRx3 dataset, we correct for batch effects using \textit{Typical Variation Normalization} (TVN) \citep{TVN2017}, and also correct for possible chromosome arm biases known to exist in CRISPR-Cas9 HCS data \citep{lazar2023high}. Notably, recall on JUMP-CP is considerably lower than on RxRx3 likely due to different assay protocols and more variance in the data.

\section{Conclusion}

We introduce \textbf{RxRx3-core}, a compact and accessible dataset with an associated set of benchmarks designed to evaluate representation learning models for high-content screening microscopy data against zero-shot drug-target and gene-gene interaction prediction. At just 18GB, RxRx3-core is available on HuggingFace and Polaris, making it practical for widespread use. Our analysis includes baseline and foundation model performance across perturbation signal magnitude and zero-shot drug-target and gene-gene interaction prediction benchmarks. The results highlight the advantages of self-supervised vision transformer models, particularly as they scale to multi-billion parameter regimes, in capturing biologically meaningful representations.
We hope that RxRx3-core and its benchmarks serve as a valuable resource for the machine learning community, fostering innovation in representation learning approaches and advancing the analysis of large-scale microscopy datasets.

\section*{Meaningfulness Statement}
A meaningful representation of life characterizes relevant aspects of the vital functions of a sample. This can be measured by benchmarking representations of experimental data to real-world known relationships. We evaluate different representation learning methods on images of HUVEC cells in our newly released public dataset, RxRx3-core, to determine how those cells are phenotypically impacted by genetic and chemical perturbations. The most advanced self-supervised model, Phenom-2 \citep{vitally_consistent}, creates the most meaningful representations in this context. We believe that this work will enable development of new methods for connecting fundamental genetic functions with compound activity for the purposes of drug discovery.

\section*{Acknowledgments}
We thank M. McKeithen, S. Moore, E. Hurst, and J. Fryer for their help computing and accessing CellProfiler features for this manuscript and the incredible Recursion lab and engineering teams for design and execution of experiments and storage and processing of data.

\bibliography{Rxrx3-core}
\bibliographystyle{iclr2025_conference}

\newpage
\appendix
\section{Appendix}

\begin{table}[H]

\centering
\begin{tabular}{p{2cm}p{11cm}}
\toprule
\textbf{Gene} & \textbf{Compounds} \\
\midrule

ACE & Moexipril, Losartan, Quinapril, Enalapril, Benazepril, Sitagliptin phosphate monohydrate, bucillamine, zofenopril-calcium \\

AR & Flutamide, Enzalutamide, Liothyronine, Indomethacin, Mifepristone, ODM-201, ARN-509, Eugenol, Nilutamide, Eplerenone, Pamoic acid disodium salt, Dihydrotestosterone(DHT), salicylamide, bicalutamide \\

BRAF & Gefitinib, Imatinib, Sorafenib, Erlotinib, Vemurafenib, Dabrafenib, BIRB 796, PD0325901, Regorafenib, RepSox, Niclosamide, CHEMBL410295, MLN 2480 \\

EGFR & Adrucil, Flavopiridol, Gefitinib, Lapatinib, BMS-777607, Phloretin, Crizotinib, BMS-536924, Quercetin, Tivozanib, Imatinib, Foretinib, Sorafenib, Erlotinib, Vemurafenib, OSI-906, Indomethacin, PD168393, Icotinib, BIRB 796, LDK378, NVP-AEW541, Pexidartinib, Ketoprofen, Vorinostat, Aspirin, AZD9291, Ibrutinib, Cabozantinib, Vandetanib, Ponatinib, LY2228820, Entrectinib, Dasatinib, Bosutinib, Niclosamide, Olmutinib, Acalabrutinib, Curcumin, brigatinib \\

ESR1 & Phloretin, Nabumetone, Raloxifene, Bazedoxifene, Letrozole, Mifepristone, Hexachlorophene, Molsidomine, Nordihydroguaiaretic acid, Resorcinol, Eplerenone \\

HDAC1 & Atorvastatin, Divalproex, Fluconazole, Bendamustine, Lapatinib, Imatinib, Celecoxib, Daunorubicin, Givinostat, Phenylbutyrate, PCI-24781, Vorinostat, AZD9291, Bortezomib, Vandetanib, LMK-235, BIIB021, Colchicine, Bestatin, Curcumin, lovastatin \\

MTOR & BGT226, GSK2126458, AZD2014, CAL-101, GDC-0349, BYL719, Dasatinib, PF-04691502, INK 128, Gedatolisib, Torin 1, LY3023414, GDC-0084, Dactolisib (BEZ235) \\

PTGS2 & Acetaminophen, Diclofenac, Naproxen, Quercetin, Vortioxetine, Celecoxib, Erlotinib, Meloxicam, Mesalamine, Indomethacin, Nimesulide, Lumiracoxib, Zileuton, Ketoprofen, Valdecoxib, Aspirin, Piroxicam, Salicylate, Eugenol, Nordihydroguaiaretic acid, Parecoxib, Loxoprofen, Curcumin, Etoricoxib, Flurizan, ibuprofen-(s), meclofenamic-acid, flurbiprofen-(S)-(+), carprofen, ketorolac, deracoxib, tepoxalin, firocoxib \\

SRC & AMG-458, Flavopiridol, Crizotinib, Quercetin, Imatinib, Nilotinib, PD168393, BIRB 796, NVP-AEW541, Enzastaurin, Ibrutinib, Cabozantinib, Vandetanib, Ponatinib, CH5183284, PRT062607, Dasatinib, Bosutinib, Mebendazole, Niclosamide, NVP-TAE 684, CHEMBL410295, AZM475271, Acalabrutinib, fosfosal \\

TP53 & BYL719 \\

\bottomrule
\end{tabular}

\caption{Known gene-compound associations used for benchmarking, showing FDA-approved drugs with their validated molecular targets. The exemplary subset includes major therapeutic targets such as kinases (EGFR, BRAF), nuclear receptors (AR, ESR1), and enzymes (ACE, PTGS2).}

\label{tab:compounds_gene}
\end{table}

\end{document}

%% file: math_commands.tex
\usepackage{amsmath,amsfonts,bm}

\def\eqref#1{equation~\ref{#1}}

\def\1{\bm{1}}

\DeclareMathAlphabet{\mathsfit}{\encodingdefault}{\sfdefault}{m}{sl}
\SetMathAlphabet{\mathsfit}{bold}{\encodingdefault}{\sfdefault}{bx}{n}













%% file: Rxrx3-core.bbl
\begin{thebibliography}{32}
\providecommand{\natexlab}[1]{#1}
\providecommand{\url}[1]{\texttt{#1}}
\expandafter\ifx\csname urlstyle\endcsname\relax
  \providecommand{\doi}[1]{doi: #1}\else
  \providecommand{\doi}{doi: \begingroup \urlstyle{rm}\Url}\fi

\bibitem[Ando et~al.(2017)Ando, McLean, and Berndl]{TVN2017}
D.~Michael Ando, Cory~Y. McLean, and Marc Berndl.
\newblock {Improving Phenotypic Measurements in High-Content Imaging Screens}.
\newblock \emph{bioRxiv}, pp.\  161422, 2017.
\newblock \doi{10.1101/161422}.

\bibitem[Arevalo et~al.(2024)Arevalo, Su, Carpenter, and Singh]{broad_motive}
John Arevalo, Ellen Su, Anne~E Carpenter, and Shantanu Singh.
\newblock {MOTI$\mathcal{VE}$: A Drug-Target Interaction Graph For Inductive Link Prediction}.
\newblock \emph{arXiv}, 2024.
\newblock \doi{10.48550/arxiv.2406.08649}.

\bibitem[Boutros et~al.(2015)Boutros, Heigwer, and Laufer]{BoutrosHCS2015}
Michael Boutros, Florian Heigwer, and Christina Laufer.
\newblock {Microscopy-Based High-Content Screening}.
\newblock \emph{Cell}, 163\penalty0 (6):\penalty0 1314--1325, 2015.
\newblock ISSN 0092-8674.
\newblock \doi{10.1016/j.cell.2015.11.007}.

\bibitem[Bray et~al.(2016)Bray, Singh, Han, Davis, Borgeson, Hartland, Kost-Alimova, Gustafsdottir, Gibson, and Carpenter]{10.1038/nprot.2016.105}
Mark-Anthony Bray, Shantanu Singh, Han Han, Chadwick~T Davis, Blake Borgeson, Cathy Hartland, Maria Kost-Alimova, Sigrun~M Gustafsdottir, Christopher~C Gibson, and Anne~E Carpenter.
\newblock {Cell Painting, a high-content image-based assay for morphological profiling using multiplexed fluorescent dyes}.
\newblock \emph{Nature Protocols}, 11\penalty0 (9):\penalty0 1757--1774, 2016.
\newblock ISSN 1754-2189.
\newblock \doi{10.1038/nprot.2016.105}.

\bibitem[Caicedo et~al.(2017)Caicedo, Cooper, Heigwer, Warchal, Qiu, Molnar, Vasilevich, Barry, Bansal, Kraus, Wawer, Paavolainen, Herrmann, Rohban, Hung, Hennig, Concannon, Smith, Clemons, Singh, Rees, Horvath, Linington, and Carpenter]{10.1038/nmeth.4397}
Juan~C Caicedo, Sam Cooper, Florian Heigwer, Scott Warchal, Peng Qiu, Csaba Molnar, Aliaksei~S Vasilevich, Joseph~D Barry, Harmanjit~Singh Bansal, Oren Kraus, Mathias Wawer, Lassi Paavolainen, Markus~D Herrmann, Mohammad Rohban, Jane Hung, Holger Hennig, John Concannon, Ian Smith, Paul~A Clemons, Shantanu Singh, Paul Rees, Peter Horvath, Roger~G Linington, and Anne~E Carpenter.
\newblock {Data-analysis strategies for image-based cell profiling}.
\newblock \emph{Nature Methods}, 14\penalty0 (9):\penalty0 849--863, 2017.
\newblock ISSN 1548-7091.
\newblock \doi{10.1038/nmeth.4397}.

\bibitem[Carpenter et~al.(2006)Carpenter, Jones, Lamprecht, Clarke, Kang, Friman, Guertin, Chang, Lindquist, Moffat, et~al.]{carpenter2006cellprofiler}
Anne~E Carpenter, Thouis~R Jones, Michael~R Lamprecht, Colin Clarke, In~Han Kang, Ola Friman, David~A Guertin, Joo~Han Chang, Robert~A Lindquist, Jason Moffat, et~al.
\newblock Cellprofiler: image analysis software for identifying and quantifying cell phenotypes.
\newblock \emph{Genome biology}, 7:\penalty0 1--11, 2006.

\bibitem[Celik et~al.(2024)Celik, Hütter, Carlos, Lazar, Mohan, Tillinghast, Biancalani, Fay, Earnshaw, and Haque]{Celik2022.12.09.519400}
Safiye Celik, Jan-Christian Hütter, Sandra~Melo Carlos, Nathan~H. Lazar, Rahul Mohan, Conor Tillinghast, Tommaso Biancalani, Marta~M. Fay, Berton~A. Earnshaw, and Imran~S. Haque.
\newblock Building, benchmarking, and exploring perturbative maps of transcriptional and morphological data.
\newblock \emph{PLOS Computational Biology}, 20\penalty0 (10):\penalty0 1--24, 10 2024.
\newblock \doi{10.1371/journal.pcbi.1012463}.
\newblock URL \url{https://doi.org/10.1371/journal.pcbi.1012463}.

\bibitem[Chandrasekaran et~al.(2023)Chandrasekaran, Ackerman, Alix, Ando, Arevalo, Bennion, Boisseau, Borowa, Boyd, Brino, et~al.]{chandrasekaran2023jump}
Srinivas~Niranj Chandrasekaran, Jeanelle Ackerman, Eric Alix, D~Michael Ando, John Arevalo, Melissa Bennion, Nicolas Boisseau, Adriana Borowa, Justin~D Boyd, Laurent Brino, et~al.
\newblock Jump cell painting dataset: morphological impact of 136,000 chemical and genetic perturbations.
\newblock \emph{bioRxiv}, pp.\  2023--03, 2023.

\bibitem[Chandrasekaran et~al.(2024)Chandrasekaran, Cimini, Goodale, Miller, Kost-Alimova, Jamali, Doench, Fritchman, Skepner, Melanson, Kalinin, Arevalo, Haghighi, Caicedo, Kuhn, Hernandez, Berstler, Shafqat-Abbasi, Root, Swalley, Garg, Singh, and Carpenter]{cpjump1}
Srinivas~Niranj Chandrasekaran, Beth~A. Cimini, Amy Goodale, Lisa Miller, Maria Kost-Alimova, Nasim Jamali, John~G. Doench, Briana Fritchman, Adam Skepner, Michelle Melanson, Alexandr~A. Kalinin, John Arevalo, Marzieh Haghighi, Juan~C. Caicedo, Daniel Kuhn, Desiree Hernandez, James Berstler, Hamdah Shafqat-Abbasi, David~E. Root, Susanne~E. Swalley, Sakshi Garg, Shantanu Singh, and Anne~E. Carpenter.
\newblock {Three million images and morphological profiles of cells treated with matched chemical and genetic perturbations}.
\newblock \emph{Nature Methods}, 21\penalty0 (6):\penalty0 1114--1121, 2024.
\newblock ISSN 1548-7091.
\newblock \doi{10.1038/s41592-024-02241-6}.

\bibitem[Drew et~al.(2017)Drew, Lee, Huizar, Tu, Borgeson, McWhite, Ma, Wallingford, and Marcotte]{hu_map}
Kevin Drew, Chanjae Lee, Ryan~L Huizar, Fan Tu, Blake Borgeson, Claire~D McWhite, Yun Ma, John~B Wallingford, and Edward~M Marcotte.
\newblock {Integration of over 9,000 mass spectrometry experiments builds a global map of human protein complexes}.
\newblock \emph{Molecular Systems Biology}, 13\penalty0 (6):\penalty0 932, 2017.
\newblock ISSN 1744-4292.
\newblock \doi{10.15252/msb.20167490}.

\bibitem[Eulenberg et~al.(2017)Eulenberg, Köhler, Blasi, Filby, Carpenter, Rees, Theis, and Wolf]{BBBC048}
Philipp Eulenberg, Niklas Köhler, Thomas Blasi, Andrew Filby, Anne~E. Carpenter, Paul Rees, Fabian~J. Theis, and F.~Alexander Wolf.
\newblock {Reconstructing cell cycle and disease progression using deep learning}.
\newblock \emph{Nature Communications}, 8\penalty0 (1):\penalty0 463, 2017.
\newblock \doi{10.1038/s41467-017-00623-3}.

\bibitem[Fay et~al.(2023)Fay, Kraus, Victors, Arumugam, Vuggumudi, Urbanik, Hansen, Celik, Cernek, Jagannathan, et~al.]{fay2023rxrx3}
Marta~M Fay, Oren Kraus, Mason Victors, Lakshmanan Arumugam, Kamal Vuggumudi, John Urbanik, Kyle Hansen, Safiye Celik, Nico Cernek, Ganesh Jagannathan, et~al.
\newblock Rxrx3: Phenomics map of biology.
\newblock \emph{bioRxiv}, pp.\  2023--02, 2023.

\bibitem[Gillespie et~al.(2021)Gillespie, Jassal, Stephan, Milacic, Rothfels, Senff-Ribeiro, Griss, Sevilla, Matthews, Gong, Deng, Varusai, Ragueneau, Haider, May, Shamovsky, Weiser, Brunson, Sanati, Beckman, Shao, Fabregat, Sidiropoulos, Murillo, Viteri, Cook, Shorser, Bader, Demir, Sander, Haw, Wu, Stein, Hermjakob, and D’Eustachio]{REACTOME}
Marc Gillespie, Bijay Jassal, Ralf Stephan, Marija Milacic, Karen Rothfels, Andrea Senff-Ribeiro, Johannes Griss, Cristoffer Sevilla, Lisa Matthews, Chuqiao Gong, Chuan Deng, Thawfeek Varusai, Eliot Ragueneau, Yusra Haider, Bruce May, Veronica Shamovsky, Joel Weiser, Timothy Brunson, Nasim Sanati, Liam Beckman, Xiang Shao, Antonio Fabregat, Konstantinos Sidiropoulos, Julieth Murillo, Guilherme Viteri, Justin Cook, Solomon Shorser, Gary Bader, Emek Demir, Chris Sander, Robin Haw, Guanming Wu, Lincoln Stein, Henning Hermjakob, and Peter D’Eustachio.
\newblock {The reactome pathway knowledgebase 2022}.
\newblock \emph{Nucleic Acids Research}, 50\penalty0 (D1):\penalty0 D687--D692, 2021.
\newblock ISSN 0305-1048.
\newblock \doi{10.1093/nar/gkab1028}.

\bibitem[Giurgiu et~al.(2019)Giurgiu, Reinhard, Brauner, Dunger-Kaltenbach, Fobo, Frishman, Montrone, and Ruepp]{CORUM}
Madalina Giurgiu, Julian Reinhard, Barbara Brauner, Irmtraud Dunger-Kaltenbach, Gisela Fobo, Goar Frishman, Corinna Montrone, and Andreas Ruepp.
\newblock {CORUM: the comprehensive resource of mammalian protein complexes—2019}.
\newblock \emph{Nucleic Acids Research}, 47\penalty0 (Database issue):\penalty0 D559--D563, 2019.
\newblock ISSN 0305-1048.
\newblock \doi{10.1093/nar/gky973}.

\bibitem[Harding et~al.(2024)Harding, Armstrong, Faccenda, Southan, Alexander, Davenport, Spedding, and Davies]{guide2pharma2024}
Simon~D Harding, Jane~F Armstrong, Elena Faccenda, Christopher Southan, Stephen P~H Alexander, Anthony~P Davenport, Michael Spedding, and Jamie~A Davies.
\newblock The iuphar/bps guide to pharmacology in 2024.
\newblock \emph{Nucleic Acids Research}, 52\penalty0 (D1):\penalty0 D1438–D1449, Jan 2024.
\newblock \doi{10.1093/nar/gkad944}.

\bibitem[He et~al.(2022)He, Chen, Xie, Li, Doll{\'a}r, and Girshick]{he2022masked}
Kaiming He, Xinlei Chen, Saining Xie, Yanghao Li, Piotr Doll{\'a}r, and Ross Girshick.
\newblock Masked autoencoders are scalable vision learners.
\newblock In \emph{Proceedings of the IEEE/CVF conference on computer vision and pattern recognition}, pp.\  16000--16009, 2022.

\bibitem[Iyer et~al.(2024)Iyer, Michael, Chi, Arevalo, Chandrasekaran, Carpenter, Rajpurkar, and Singh]{CPJUMP1_supervised}
Niveditha~S. Iyer, Daniel~J. Michael, S-Y~Gordon Chi, John Arevalo, Srinivas~Niranj Chandrasekaran, Anne~E. Carpenter, Pranav Rajpurkar, and Shantanu Singh.
\newblock {Cell morphological representations of genes enhance prediction of drug targets}.
\newblock \emph{bioRxiv}, pp.\  2024.06.08.598076, 2024.
\newblock \doi{10.1101/2024.06.08.598076}.

\bibitem[Kamentsky et~al.(2011)Kamentsky, Jones, Fraser, Bray, Logan, Madden, Ljosa, Rueden, Eliceiri, and Carpenter]{kamentsky2011improved}
Lee Kamentsky, Thouis~R Jones, Adam Fraser, Mark-Anthony Bray, David~J Logan, Katherine~L Madden, Vebjorn Ljosa, Curtis Rueden, Kevin~W Eliceiri, and Anne~E Carpenter.
\newblock Improved structure, function and compatibility for cellprofiler: modular high-throughput image analysis software.
\newblock \emph{Bioinformatics}, 27\penalty0 (8):\penalty0 1179--1180, 2011.

\bibitem[Kenyon-Dean et~al.(2024)Kenyon-Dean, Wang, Urbanik, Donhauser, Hartford, Saberian, Sahin, Bendidi, Celik, Fay, Vera, Haque, and Kraus]{vitally_consistent}
Kian Kenyon-Dean, Zitong~Jerry Wang, John Urbanik, Konstantin Donhauser, Jason Hartford, Saber Saberian, Nil Sahin, Ihab Bendidi, Safiye Celik, Marta Fay, Juan Sebastian~Rodriguez Vera, Imran~S Haque, and Oren Kraus.
\newblock {ViTally Consistent: Scaling Biological Representation Learning for Cell Microscopy}.
\newblock In \emph{38th Conference on Neural Information Processing Systems (NeurIPS) - Workshop on Foundation Models for Science: Progress, Opportunities, and Challenges}, 2024.
\newblock URL \url{https://arxiv.org/abs/2411.02572}.

\bibitem[Kraus et~al.(2024)Kraus, Kenyon-Dean, Saberian, Fallah, McLean, Leung, Sharma, Khan, Balakrishnan, Celik, et~al.]{kraus2024masked}
Oren Kraus, Kian Kenyon-Dean, Saber Saberian, Maryam Fallah, Peter McLean, Jess Leung, Vasudev Sharma, Ayla Khan, Jia Balakrishnan, Safiye Celik, et~al.
\newblock Masked autoencoders for microscopy are scalable learners of cellular biology.
\newblock In \emph{Proceedings of the IEEE/CVF Conference on Computer Vision and Pattern Recognition}, pp.\  11757--11768, 2024.

\bibitem[Lazar et~al.(2024)Lazar, Celik, Chen, Fay, Irish, Jensen, Tillinghast, Urbanik, Bone, Gibson, et~al.]{lazar2023high}
Nathan~H Lazar, Safiye Celik, Lu~Chen, Marta~M Fay, Jonathan~C Irish, James Jensen, Conor~A Tillinghast, John Urbanik, William~P Bone, Christopher~C Gibson, et~al.
\newblock High-resolution genome-wide mapping of chromosome-arm-scale truncations induced by crispr--cas9 editing.
\newblock \emph{Nature Genetics}, pp.\  1--12, 2024.

\bibitem[Liu et~al.(2007)Liu, Lin, Wen, Jorissen, and Gilson]{bindingdb2007}
Tiqing Liu, Yuhmei Lin, Xin Wen, Robert~N Jorissen, and Michael~K Gilson.
\newblock Bindingdb: a web-accessible database of experimentally determined protein-ligand binding affinities.
\newblock \emph{Nucleic Acids Research}, 35\penalty0 (Database issue):\penalty0 D198--201, Jan 2007.
\newblock \doi{10.1093/nar/gkl999}.

\bibitem[Ljosa et~al.(2013)Ljosa, Caie, Horst, Sokolnicki, Jenkins, Daya, Roberts, Jones, Singh, Genovesio, Clemons, Carragher, and Carpenter]{BBBC021}
Vebjorn Ljosa, Peter~D. Caie, Rob~ter Horst, Katherine~L. Sokolnicki, Emma~L. Jenkins, Sandeep Daya, Mark~E. Roberts, Thouis~R. Jones, Shantanu Singh, Auguste Genovesio, Paul~A. Clemons, Neil~O. Carragher, and Anne~E. Carpenter.
\newblock {Comparison of Methods for Image-Based Profiling of Cellular Morphological Responses to Small-Molecule Treatment}.
\newblock \emph{SLAS Discovery}, 18\penalty0 (10):\penalty0 1321--1329, 2013.
\newblock ISSN 2472-5552.
\newblock \doi{10.1177/1087057113503553}.

\bibitem[Ouyang et~al.(2019)Ouyang, Winsnes, Hjelmare, Cesnik, Åkesson, Xu, Sullivan, Dai, Lan, Jinmo, Galib, Henkel, Hwang, Poplavskiy, Tunguz, Wolfinger, Gu, Li, Xie, Buslov, Fironov, Kiselev, Panchenko, Cao, Wei, Wu, Zhu, Tseng, Gao, Ju, Yi, Zheng, Kappel, and Lundberg]{hpa_localization}
Wei Ouyang, Casper~F. Winsnes, Martin Hjelmare, Anthony~J. Cesnik, Lovisa Åkesson, Hao Xu, Devin~P. Sullivan, Shubin Dai, Jun Lan, Park Jinmo, Shaikat~M. Galib, Christof Henkel, Kevin Hwang, Dmytro Poplavskiy, Bojan Tunguz, Russel~D. Wolfinger, Yinzheng Gu, Chuanpeng Li, Jinbin Xie, Dmitry Buslov, Sergei Fironov, Alexander Kiselev, Dmytro Panchenko, Xuan Cao, Runmin Wei, Yuanhao Wu, Xun Zhu, Kuan-Lun Tseng, Zhifeng Gao, Cheng Ju, Xiaohan Yi, Hongdong Zheng, Constantin Kappel, and Emma Lundberg.
\newblock {Analysis of the Human Protein Atlas Image Classification competition}.
\newblock \emph{Nature Methods}, 16\penalty0 (12):\penalty0 1254--1261, 2019.
\newblock ISSN 1548-7091.
\newblock \doi{10.1038/s41592-019-0658-6}.

\bibitem[Przybyla \& Gilbert(2022)Przybyla and Gilbert]{PrzybylaNatGeneticsReviews2022}
Laralynne Przybyla and Luke~A. Gilbert.
\newblock {A new era in functional genomics screens}.
\newblock \emph{Nature Reviews Genetics}, 23\penalty0 (2):\penalty0 89--103, 2022.
\newblock ISSN 1471-0056.
\newblock \doi{10.1038/s41576-021-00409-w}.

\bibitem[Rizzo \& Székely(2016)Rizzo and Székely]{energy_distance}
Maria~L. Rizzo and Gábor~J. Székely.
\newblock {Energy distance}.
\newblock \emph{Wiley Interdisciplinary Reviews: Computational Statistics}, 8\penalty0 (1):\penalty0 27--38, 2016.
\newblock ISSN 1939-5108.
\newblock \doi{10.1002/wics.1375}.

\bibitem[Stirling et~al.(2021)Stirling, Swain-Bowden, Lucas, Carpenter, Cimini, and Goodman]{10.1186/s12859-021-04344-9}
David~R. Stirling, Madison~J. Swain-Bowden, Alice~M. Lucas, Anne~E. Carpenter, Beth~A. Cimini, and Allen Goodman.
\newblock {CellProfiler 4: improvements in speed, utility and usability}.
\newblock \emph{BMC Bioinformatics}, 22\penalty0 (1):\penalty0 433, 2021.
\newblock \doi{10.1186/s12859-021-04344-9}.

\bibitem[Szklarczyk et~al.(2020)Szklarczyk, Gable, Nastou, Lyon, Kirsch, Pyysalo, Doncheva, Legeay, Fang, Bork, Jensen, and Mering]{stringdb}
Damian Szklarczyk, Annika~L Gable, Katerina~C Nastou, David Lyon, Rebecca Kirsch, Sampo Pyysalo, Nadezhda~T Doncheva, Marc Legeay, Tao Fang, Peer Bork, Lars~J Jensen, and Christian~von Mering.
\newblock {The STRING database in 2021: customizable protein–protein networks, and functional characterization of user-uploaded gene/measurement sets}.
\newblock \emph{Nucleic Acids Research}, 49\penalty0 (D1):\penalty0 D605--D612, 2020.
\newblock ISSN 0305-1048.
\newblock \doi{10.1093/nar/gkaa1074}.

\bibitem[Vincent et~al.(2022)Vincent, Nueda, Lee, Schenone, Prunotto, and Mercola]{VincentPhenotypic2022}
Fabien Vincent, Arsenio Nueda, Jonathan Lee, Monica Schenone, Marco Prunotto, and Mark Mercola.
\newblock {Phenotypic drug discovery: recent successes, lessons learned and new directions}.
\newblock \emph{Nature Reviews Drug Discovery}, 21\penalty0 (12):\penalty0 899--914, 2022.
\newblock ISSN 1474-1776.
\newblock \doi{10.1038/s41573-022-00472-w}.

\bibitem[Wognum et~al.(2024)Wognum, Ash, Aldeghi, Rodríguez-Pérez, Fang, Cheng, Price, Clevert, Engkvist, and Walters]{polaris_NMI}
Cas Wognum, Jeremy~R. Ash, Matteo Aldeghi, Raquel Rodríguez-Pérez, Cheng Fang, Alan~C. Cheng, Daniel~J. Price, Djork-Arné Clevert, Ola Engkvist, and W.~Patrick Walters.
\newblock {A call for an industry-led initiative to critically assess machine learning for real-world drug discovery}.
\newblock \emph{Nature Machine Intelligence}, 6\penalty0 (10):\penalty0 1120--1121, 2024.
\newblock \doi{10.1038/s42256-024-00911-w}.

\bibitem[Wognum et~al.(2025)Wognum, Zhu, Mary, St-Laurent, Larissa, Quirke, Hounwanou, Zhu, Whitfield, Burns, (McLean), (McLean), and felix]{polaris_zenodo}
Cas Wognum, Lu~Zhu, Hadrien Mary, Julien St-Laurent, Larissa, Andrew Quirke, Honoré Hounwanou, Kun Zhu, Shawn Whitfield, Jackson Burns, Kira~Howe (McLean), Kira~Howe (McLean), and felix.
\newblock polaris-hub/polaris: 0.11.8, February 2025.
\newblock URL \url{https://doi.org/10.5281/zenodo.14834887}.

\bibitem[Zdrazil et~al.(2024)Zdrazil, Felix, Hunter, Manners, Blackshaw, Corbett, de~Veij, Ioannidis, Lopez, Mosquera, and et~al.]{chembl2024}
Barbara Zdrazil, Eloy Felix, Fiona Hunter, Emma~J Manners, James Blackshaw, Sybilla Corbett, Marleen de~Veij, Harris Ioannidis, David~Mendez Lopez, Juan~F Mosquera, and et~al.
\newblock The chembl database in 2023: a drug discovery platform spanning multiple bioactivity data types and time periods.
\newblock \emph{Nucleic Acids Research}, 52\penalty0 (D1):\penalty0 D1180–D1192, Jan 2024.
\newblock \doi{10.1093/nar/gkad1004}.

\end{thebibliography}
